\documentclass[aps,prl,reprint,showpacs,superscriptaddress,groupedaddress]{revtex4-1}
\usepackage{graphicx}
\usepackage{hyperref}
\usepackage{dcolumn}   
\usepackage{bm}        
\usepackage{amssymb}   
\newcommand{\ket}[1]{{\left| {#1} \right\rangle}}
\newcommand{\bra}[1]{{\left\langle {#1} \right|}}
\newcommand{\ketbra}[2]{{\left| {#1} \right\rangle \!\!\left\langle {#2} \right|}}
\begin{document}

\title{Direct frequency comb laser cooling and trapping}
\author{A. M. Jayich}
\email{jayich@gmail.com}
\author{ X. Long}
\author{W. C. Campbell}
\affiliation{UCLA Department of Physics and Astronomy, Los Angeles, California 90095, USA}
\affiliation{California Institute for Quantum Emulation, Santa Barbara, California 93106, USA}

\date{\today}

\begin{abstract}
Continuous wave (CW) lasers are the enabling technology for producing
ultracold atoms and molecules through laser cooling and trapping.  The
resulting pristine samples of slow moving particles are the de facto
starting point for both fundamental and applied science when a
highly-controlled quantum system is required. Laser cooled atoms have
recently led to major advances in quantum
information \cite{Garcia-Ripoll:2005nx,Bloch:2012fk}, the search to
understand dark energy \cite{Hamilton:2015yq}, quantum
chemistry \cite{Zhu:2014fj,Miranda:2011cr}, and quantum
sensors \cite{Dickerson:2013rc}. However, CW laser technology currently
limits laser cooling and trapping to special types of elements that do
not include highly abundant and chemically relevant atoms such as
hydrogen, carbon, oxygen, and nitrogen. Here, we demonstrate that
Doppler cooling and trapping by optical frequency combs may provide a
route to trapped, ultracold atoms whose spectra are not amenable to CW
lasers. We laser cool a gas of atoms by driving a two-photon
transition with an optical frequency
comb \cite{Kielpinski:2006ai,MarianSCIENCE04}, an efficient process to
which every comb tooth coherently contributes \cite{BaklanovAP77}. We
extend this technique to create a magneto-optical trap (MOT), an
electromagnetic beaker for accumulating the laser-cooled atoms for
further study. Our results suggest that the efficient frequency
conversion offered by optical frequency combs could provide a key
ingredient for producing trapped, ultracold samples of nature's most
abundant building blocks, as well as antihydrogen. As such, the
techniques demonstrated here may enable advances in fields as
disparate as molecular biology and the search for physics beyond the
standard model.

\end{abstract}

\maketitle

High precision physical measurements are often undertaken close to
absolute zero temperature to minimize thermal fluctuations. For
example, the measurable properties of a room temperature chemical
reaction (rate, product branching, etc.) include a thermally-induced
average over a large number of reactant and product quantum states,
which masks the unique details of specific reactant-product
pairs. Doppler laser cooling with CW lasers is a
robust method to reduce the random motion of atoms \cite{WinelandPRL78}
and molecules \cite{ShumanNATURE467}. With some added complexity, the
same laser light can be made to spatially confine the atoms in a 
MOT \cite{Raab1987,HummonPRL13}. The resulting
sub-kelvin gas-phase atoms can then be studied and controlled with
high precision. This technique has begun to be applied to chemistry,
where it has recently enabled the measurement and control ultracold
chemical reactions at a new level of
detail \cite{Zhu:2014fj,Miranda:2011cr} using species made from alkali
atoms, which are well-suited to CW laser cooling and trapping.

While the prospects of comprehensive precision spectroscopy and pure
state resolution of arbitrary chemical reactions is enticing, Doppler
cooling is limited by the availability of CW lasers to a subset of
atoms and molecules that have convenient internal structure.  In
particular, the lack of sufficiently powerful CW lasers in the deep
ultraviolet (UV) means that laser cooling and trapping is not
currently available for the most prevalent atoms in organic chemistry
and living organisms: hydrogen, carbon, oxygen, and nitrogen.  Due to
their simplicity and abundance, these species likewise play prominent
roles in other scientific fields such as
astrophysics \cite{HerbstARPC95} and precision
measurement \cite{BluhmPRL99}, where the production of cold samples
could help answer fundamental outstanding questions \cite{DuttaMNRAS15,
  DonnanJPB13, HamiltonPRL14}.

\begin{figure}
\begin{center}
\includegraphics[scale=0.3]{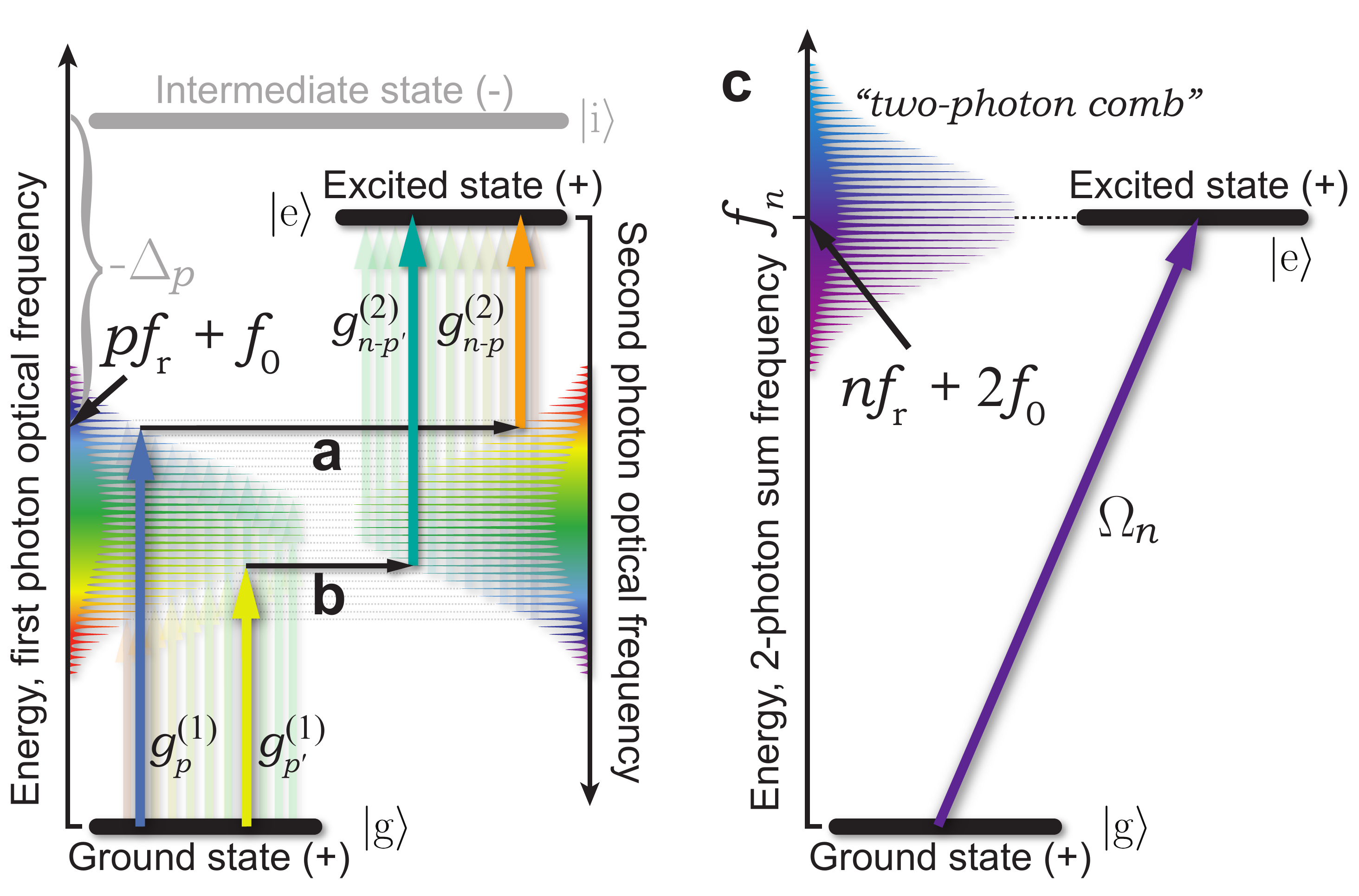}
\end{center}
\caption{Constructive interference of multiple paths in a two-photon
  transition driven by a transform-limited optical frequency comb.
  All pairs of comb teeth whose sum frequency matches the excited
  state energy interfere constructively to excite atoms.  Two example
  pairs are shown as \textbf{a} and \textbf{b}, and the effective
  two-level system that results from the sum is shown in
  \textbf{c}. Every tooth of the resulting ``two-photon comb'' of
  resonant coupling strength $\Omega$ leverages the full power of all
  of the optical frequency comb teeth through this massively-parallel
  constructive interference.}
\label{fig:Schematic}
\end{figure}

\begin{figure*}
\begin{center}
\includegraphics[scale=0.8]{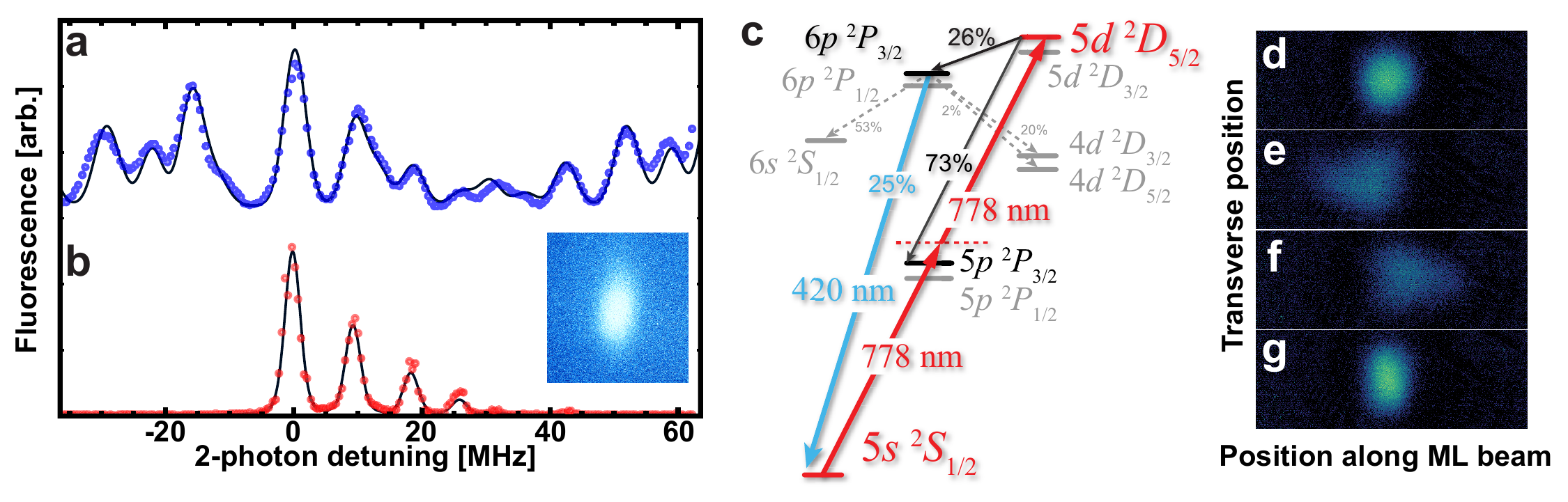}
\end{center}
\caption{Laser-induced fluorescence spectrum of the
  $5S\!\!\rightarrow\!\!5D$ two-photon transition driven by an optical
  frequency comb.  \textbf{a}: Spectrum from a natural abundance vapor
  cell and \textbf{b}: a $^{85}\mathrm{Rb}$ CW MOT and 
collected $420 \mbox{ nm}$ light
  (inset).  Solid curves are theory fitted for \textbf{a} Gaussian and
  \textbf{b} Voigt line shapes. \textbf{c}: Relevant levels of
  rubidium. \textbf{d}-\textbf{g}: absorption images of the atom cloud after free expansion
  following \textbf{d}: no ML illumination, \textbf{e}: ML
  illumination from the right, \textbf{f}: left and \textbf{g}: both directions, detuned
  to the red of resonance.  Mechanical forces are evident in
  \textbf{e} and \textbf{f}, and the narrowing of the velocity
  distribution in the horizontal direction in \textbf{g} is the
  hallmark of cooling.}
\label{fig:spectrum}
\end{figure*}

In contrast to CW lasers, mode-locked (ML) lasers have very high
instantaneous intensity and can therefore be efficiently frequency
multiplied to the UV. However, the spectrum of a ML laser consists of
many evenly spaced spectral lines (an optical frequency comb) spanning
a bandwidth much larger than a typical Doppler shift, and ML lasers
have therefore found very little use as control tools for cooling the
motion of atoms and molecules.  Doppler cooling with combs has been
investigated in a mode where each atom interacts with only one or two
comb teeth at a time, which uses only a small fraction of the laser's
total power \cite{StrohmeierOC89,WatanabeJOSAB96,IlinovaPRA11,AumilerPRA12, HanschPRL2016}.
Here, following the observation of a pushing force by Marian
\textit{et al.} \cite{MarianSCIENCE04} and a proposal by
Kielpinski \cite{Kielpinski:2006ai}, we utilize a coherent effect in
far-detuned ML two-photon transitions \cite{BaklanovAP77} to laser cool
atoms with all of the comb teeth contributing in parallel to enhance
the scattering rate (Fig.~\ref{fig:Schematic}). This technique is
designed to allow us to utilize the high UV conversion efficiency of
ML lasers without wasting any of the resulting UV power, and opens the
door to laser cool H, C, N, O, and anti-hydrogen ($\bar{\mbox{H}}$),
species for which single-photon laser cooling is beyond the reach of
even ML lasers \cite{Kielpinski:2006ai}. We extend these ideas to
create a magneto-optical trap, and find that the density of the comb
spectrum introduces no measurable effects in our system, demonstrating
that it may be possible to create MOTs of these species using this
technique.

A simple model can be used to describe the interaction between
three-level atoms and an optical frequency comb for two-photon laser
cooling and trapping (see Fig.~\ref{fig:Schematic} and Methods for
details). The two-photon coupling strength between the ground and
excited states in this case will also be a comb (the
\textit{``two-photon comb''} shown in
Fig.~\ref{fig:Schematic}\textbf{c}), the $n^{\mathrm{th}}$ tooth of
which is associated with a frequency $f_n = n f_{\mathrm{r}} + 2f_0$,
where $f_0$ is the carrier-envelope offset frequency of the optical
comb and $f_{\mathrm{r}}\equiv 1/T_{\mathrm{r}}$ is the pulse
repetition rate.  For a transform-limited ML laser, we can model the
effective (time-averaged) resonant Rabi frequency of the
$n^{\mathrm{th}}$ tooth of this two-photon comb as
\begin{equation}
\Omega_n = \sum_p \frac{g^{(1)}_p g^{(2)}_{n-p}}{2\Delta_p}\label{Omegan}
\end{equation}
where $g^{(1)}_p$ is the resonant single-photon Rabi frequency for
excitation from the ground $\ket{\mathrm{g}}$ to the intermediate
state $\ket{\mathrm{i}}$ due to the $p^{\mathrm{th}}$ optical comb
tooth and $g^{(2)}_p$ is the same quantity for excitation from the
intermediate state $\ket{\mathrm{i}}$ to the excited state
$\ket{\mathrm{e}}$ (Fig.~\ref{fig:Schematic}\textbf{a},\textbf{b}).
The single-photon detuning from the intermediate state is $\Delta_p =
pf_{\mathrm{r}} + f_0 - f_{\mathrm{gi}}$ where $f_{\mathrm{gi}}$ is
the intermediate state's energy divided by Planck's constant $h$ (we
take the energy of the ground state to be zero).  If we denote by $N$
the index of the two-photon comb tooth closest to resonance
(associated with the optical sum frequency $f_N = Nf_{\mathrm{r}} + 2
f_0$) and the pulse duration is short compared to the excited state's
lifetime ($\tau \equiv 1/\gamma$), we can approximate the resonant
Rabi frequency of each two-photon comb tooth in the vicinity of
resonance as being given by $\Omega_N$. In the limit of weak
single-pulse excitation ($\Omega_NT_{\mathrm{r}} \ll \pi$), the
time-averaged excitation rate for an atom moving with velocity
$\mbox{\boldmath$v$}$ is given by (see
Ref.~ \cite{IlinovaPRA11,AumilerPRA12} and Methods)
\begin{equation}
\gamma_{\mathrm{comb}} = 
\frac{\Omega_N^2 T_{\mathrm{r}}}{4}
\frac{ \mathrm{sinh}\!\left(
  \gamma \,T_{\mathrm{r}}/2 \right)}{ \mathrm{cosh}\!\left(
  \gamma \,T_{\mathrm{r}}/2 \right) -\cos\! \left( \delta_N(\mbox{\boldmath$v$})\,
  T_{\mathrm{r}} \right)} \label{sinhcosh}
\end{equation}
where $\delta_N(\mbox{\boldmath$v$}) \equiv 2 \pi(f_N -
f_{\mathrm{ge}} - f_N \mbox{\boldmath$\hat{k}$}\cdot
\mbox{\boldmath$v$}/c)$ is the detuning of the $N^{\mathrm{th}}$
two-photon comb tooth from two-photon resonance,
\boldmath$\hat{k}\mbox{ }$\unboldmath is a unit vector pointing in the
direction of laser propagation, and $f_{\mathrm{ge}}$ is the energy of
the excited state divided by $h$.  If both the detuning
$\delta_N(\mbox{\boldmath$v$})$ and natural linewidth $\gamma$ are
small compared to the comb tooth spacing ($2 \pi f_{\mathrm{r}}$),
this two-photon comb can be treated as having only a single tooth
(monochromatic interaction) with a two-photon Rabi frequency of
$\Omega_N$. Most of the work we describe here takes place once the
atoms are fairly cold ($kv \ll 2 \pi f_{\mathrm{r}}$) in this ``single
two-photon tooth limit,'' which gives rise to an excitation
rate of
\begin{equation}
\gamma_{N} = \frac{\Omega_N^2}{\gamma} \frac{1}{1 +
  \left(2\delta_N(\mbox{\boldmath$v$})/\gamma \right)^2}.\label{SingleToothExcitationRate}
\end{equation} 
Since the AC Stark shifts from the proximity of the intermediate state
to the optical photon energy are the same order of magnitude as
$\Omega_N$, they can be neglected compared to the linewidth in the
low-saturation limit.  For cases where a single laser photon has
enough energy to photoionize an excited atom, since both the
time-averaged excitation rate and the time-averaged photoionization
rate from the excited state depend only upon the time-averaged
intensity, the average ionization rate is exactly the same as for a CW
laser with the same frequency and time-averaged
power \cite{Kielpinski:2006ai}.

Using this simplification, an algebraic model for Doppler cooling can
be constructed for the degenerate two-photon case (as opposed to
two-color excitation \cite{WuPRL09}) to estimate the Doppler
temperature.  We assume that the laser's center frequency is near
$f_{\mathrm{ge}}/2$ and that the single tooth of interest in the
two-photon comb can be characterized by a two-photon saturation
parameter $s_N \!\equiv \!2 \Omega_N^2/\gamma^2 \!\ll\! 1$.  For slow
atoms ($kv \!\ll\! \gamma$), the cooling power of a 1D, two-photon
optical molasses detuned $\gamma/2$ to the red side of two-photon
resonance is given by the same expression as the single-photon CW
laser cooling case, $\partial E/\partial t |_{\mathrm{cool}} = -s_N
\hbar \omega_{\mathrm{ge}}^2 v^2/c^2$, where $\omega_{\mathrm{ge}}
\equiv 2 \pi f_{\mathrm{ge}}$.  The heating caused by momentum kicks
from absorption is likewise identical to the CW single-photon
expression, $\partial E/\partial t |_{\mathrm{heat,abs.}} = s_N \gamma
\hbar^2 \omega_{\mathrm{ge}}^2/4 m c^2$.

The heating caused by spontaneous emission, however, is modified by
both the multi-photon nature of the emission and the details of
excitation by a comb as follows.  First, the decay of the excited
state is likely to take place in multiple steps due to the parity
selection rule, splitting the de-excitation into smaller momentum
kicks that are unlikely to occur in the same direction, reducing the
heating.  Second, two-photon laser cooling with counter-propagating CW
laser beams adds heating in the form of Doppler-free (two-beam)
excitations \cite{ZehnlePRA01}, which produce no cooling force in 1D
but do cause heating through the subsequent spontaneous emission.  By
using a comb, however, one can easily eliminate these Doppler-free
transitions through timing by ensuring that pulses propagating in
different directions do not hit the atoms simultaneously.  In the
frequency domain, this delay produces a frequency-dependent phase
shift of the frequency comb for the second photon (shown on the right
side of Fig.~\ref{fig:Schematic}\textbf{a}, \textbf{b}), destroying the
coherent addition of comb teeth pairs necessary to drive the
transition.  The net result is that the heating rate from spontaneous
emission for two-photon laser cooling with an optical frequency comb
can be modeled by
\begin{equation}
\left. \frac{\partial E}{\partial t} \right|_{\mathrm{heat,spon.}} =
s_N \gamma \frac{\hbar^2 \omega_{\mathrm{ge}}^2}{8 m c^2}.\label{SponHeatingPower}
\end{equation} 
The balance between the cooling power and the
sum of these heating powers occurs at the Doppler temperature for
two-photon laser cooling with an optical frequency comb:
\begin{equation}
T_{\mathrm{D}} = \frac{3}{4} \frac{\hbar \gamma}{2
  k_{\mathrm{B}}}\label{DopplerPrediction}
\end{equation}
where $k_{\mathrm{B}}$ is the Boltzmann constant.

As a first experimental test of direct frequency comb 2-photon cooling
and trapping, we report a demonstration of the technique using
rubidium atoms.  For the $5{}^2S_{1/2}\!\!\rightarrow
\!\!5{}^2D_{5/2}$ transition in rubidium, the natural decay rate of
the excited state is $\gamma/2 \pi = 667 \mbox{
  kHz}$ \cite{ShengPRA08}.  Eq.~(\ref{DopplerPrediction}) gives a Doppler
cooling limit of $12 \mbox{ }\mu\mbox{K}$, which will also be true in
3D for a ML laser with non-colliding pulses.  In this work, we apply
cooling in 1D with spontaneous emission into 3D, and our effective
transition linewidth must also be taken into account (see Methods),
which yields a predicted Doppler limit of $31 \mbox{ }\mu\mbox{K}$ for
this system, considerably colder than the single-photon $5{}^2S_{1/2}
\!\rightarrow \!  5{}^2P_{3/2}$ 3D Doppler limit of $146 \mbox{
}\mu\mbox{K}$.

The optical frequency comb in this work is generated from a
Ti:Sapphire laser emitting $2\!\!-\!\!5 \mbox{ ps}$ pulses (less than
$500 \mbox{ GHz}$ bandwidth) at $778 \mbox{ nm}$ at a repetition rate of
$f_{\mathrm{r}}=81.14 \mbox{ MHz}$.

\begin{figure}
\begin{center}
\includegraphics[scale=0.38]{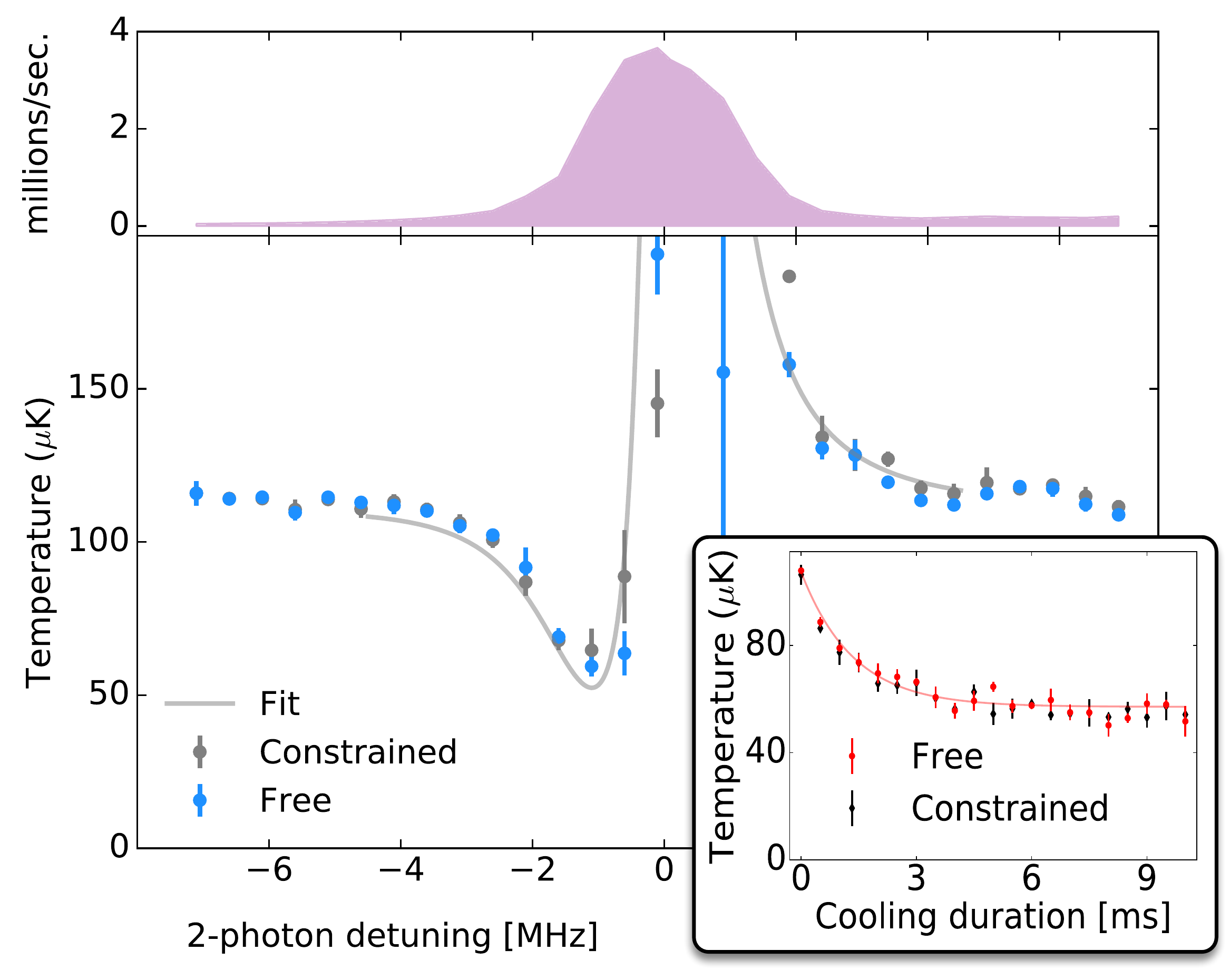}
\end{center}
\caption{Detuning dependence of $420 \mbox{ nm}$
  fluorescence (top) and the resulting temperature (bottom) of
  rubidium atoms laser-cooled by an optical frequency comb on a
  two-photon transition.  The solid curve is fit for scattering
  rate, effective linewidth and detuning offset of data
  analyzed with the aid of a monte carlo technique (data labeled
  ``Constrained,'' see Methods).  The same data are also analyzed
  using a free expansion model (``Free''), and agree well with the
  monte carlo assisted analysis. (Inset) temperature vs. time when 
  the laser detuning is optimized for cooling, giving
  a minimum temperature of $(57 \pm 2) \mbox{ }\mu\mbox{K}$. 
  Error bars are statistical over repeated measurements.}
\label{fig:molasses}
\end{figure}

We prepare an initial sample of $\approx 10^7$ $^{85}\mathrm{Rb}$
atoms using a standard CW laser MOT at $780 \mbox{ nm}$. The magnetic
field and the CW laser cooling light are then turned off, leaving the
atoms at a temperature typically near $110 \mbox{ }\mu\mbox{K}$. A
weak CW ``repump'' laser is left on continuously to optically pump
atoms out of the $F_{\mathrm{g}}=2$ ground state, and has no
measurable direct mechanical effect.  Each ML beam typically has a
time-averaged power of $(500 \pm 50) \mbox{ mW}$ and a diameter of
$(1.1 \pm 0.1) \mbox{ mm}$.  After illumination by the ML laser, the
atoms are allowed to freely expand and are subsequently imaged using
resonant CW absorption to determine their position and velocity
distributions.

By monitoring the momentum transfer from a single ML beam
(Fig.~\ref{fig:spectrum}\textbf{e}, \textbf{f}), we measure a resonant
excitation rate to be $\gamma_{\mathrm{scatt}} = (6500 \pm 700)\mbox{
  s}^{-1}$. Our theoretical estimate from Eq.~(\ref{Omegan}) and our
laser parameters gives $(13000 \pm 2000)\mbox{ s}^{-1}$, which
suggests that there may be residual chirp in the pulses that is
suppressing the excitation rate by about a factor of 2.  The measured
rate is well above the threshold needed to support these atoms against
gravity ($\approx 810 \mbox{ s}^{-1}$), which suggests that 3D
trapping should be possible with additional laser power for the
inclusion of four more beams.

We observe Doppler cooling and its dependence on two-photon detuning
by applying counter-propagating linearly-polarized ML beams to the
atom cloud for $4 \mbox{ ms}$ in zero magnetic field. By fitting the
spatial distribution of the atoms (see Methods), we extract a 1D
temperature, shown in Fig.~\ref{fig:molasses}.  The solid curve is
based on the algebraic model used above to derive the Doppler limit
and is fit for a resonant single-beam excitation rate of $(4800 \pm
400) \mbox{ s}^{-1}$ and linewidth $\gamma_{\mathrm{eff}}/2 \pi =
(1.88 \pm 0.07) \mbox{ MHz}$, consistent with the single-beam recoil
measurements.  We realize a minimum temperature of $(57 \pm 2) \mbox{
}\mu\mbox{K}$ (Fig.~\ref{fig:molasses}, inset). However, the reduced temperature is hotter
than the expected Doppler limit of $31 \mbox{
}\mu\mbox{K}$ for our system (see Methods).  We find experimentally
that the temperature inferred from free-expansion imaging is highly
sensitive to beam alignment, and therefore suspect the discrepancy is
due to imperfect balancing of the forward and backward scattering
forces at some locations in the sample \cite{Lett:1989ve}.

\begin{figure}
\begin{center}
\includegraphics[scale=0.31]{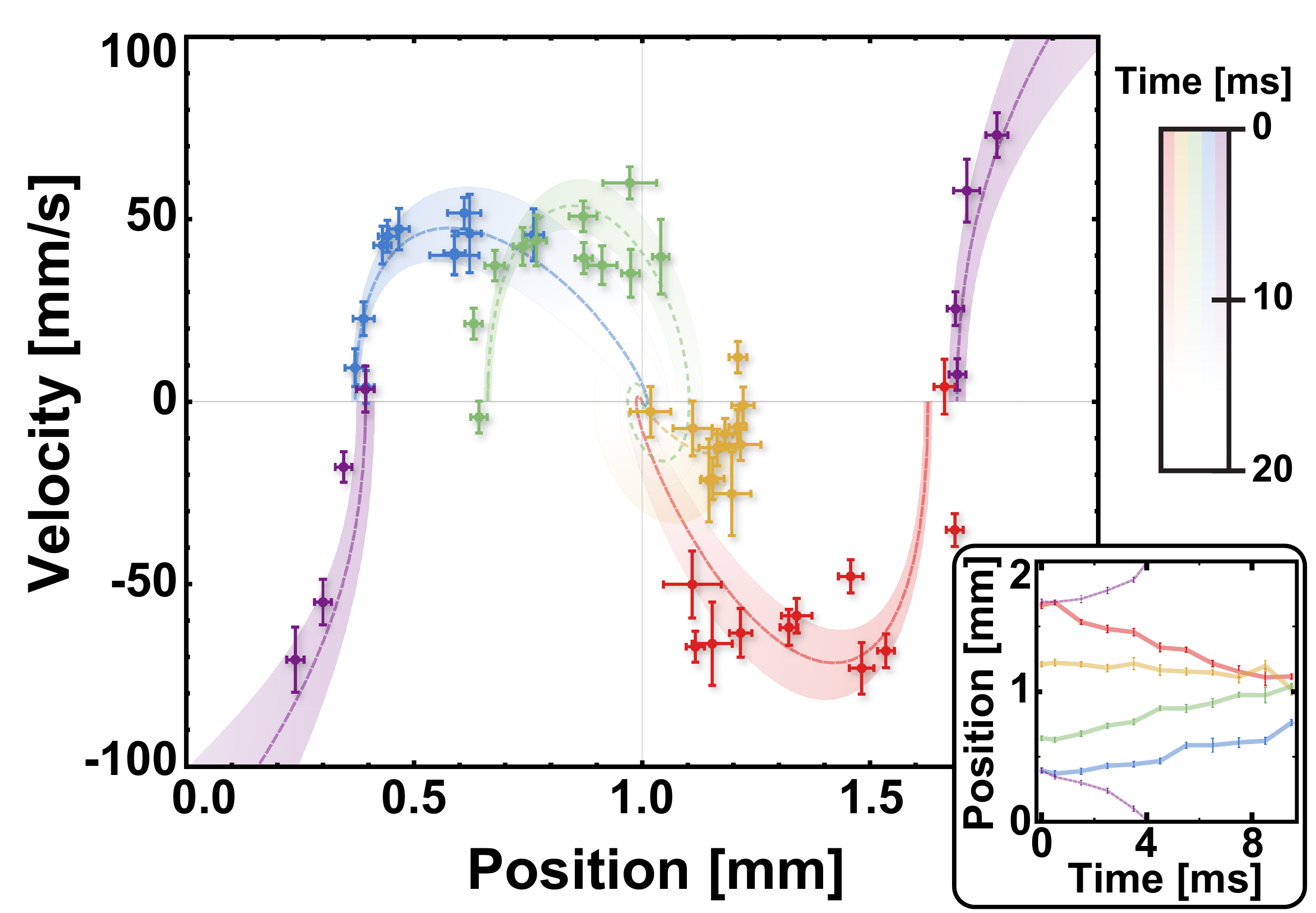}
\end{center}
\caption{Phase space (and position space, inset) trajectories for
  atoms trapped in a two-photon, optical frequency comb MOT.  Smooth
  curves are fits to a damped harmonic oscillator, with fit
  uncertainties shown as bands.  Purple features show the behavior
  when the ML beam polarizations are intentionally reversed and
  exhibit an anti-confining force. Error bars are statistical over 
  repeated measurements.}
\label{fig:PhaseSpace}
\end{figure}

To investigate the feasibility of using this technique to make a MOT,
a quadrupole magnetic field with a gradient of $7.7 \mbox{
  G}/\mbox{cm}$ is introduced and the ML beam polarizations are set to
drive $\sigma^\pm$ transitions in the standard single-photon CW MOT
configuration \cite{Raab1987}.  We displace the atom cloud from the trap center and monitor the atoms as they are pushed toward
the trap center, as shown in Fig.~\ref{fig:PhaseSpace}. The system is
modeled as a damped harmonic oscillator and fitting the motion of the
atoms yields a trapping frequency of $\nu_{\mathrm{MOT}} = (40 \pm 9)
\mbox{ Hz}$ and a cyclic damping rate of $(37 \pm 4) \mbox{ Hz}$.
These MOT parameters imply a resonant excitation rate of $(7000 \pm
1000) \mbox{ s}^{-1}$ and an effective magnetic line shift of $(0.5
\pm 0.2) \mbox{ }\mu_{\mathrm{B}}$.  The average of the calculated
line shifts for all $\Delta m_F=+2$ transitions would be $1.2 \mbox{
}\mu_{\mathrm{B}}$ for $F_{\mathrm{g}}=3 \!\! \rightarrow \!\!
F_{\mathrm{e}} = 5$, suggesting that some $\pi$-polarization may be
playing a role.  We are unable to detect any measurable
atom loss in the few milliseconds of ML illumination before atoms exit
the interaction volume due to transverse motion, consistent with the
measured photoionization cross section \cite{DuncanPRA01}.

This demonstration with rubidium shows that it may be possible to
apply these techniques in the deep UV to laser cool and
magneto-optically trap species such as H, C, N, O, and
$\bar{\mbox{H}}$ (anti-hydrogen).  Due to low anticipated scattering
rates, these species will likely need to be slowed using other
means \cite{HoganPRL08,HummonPRA08}.  Direct comb laser cooling and
trapping would then be used to cool them to the Doppler limit in a
MOT.

For H and $\bar{\mbox{H}}$, to minimize photoionization
losses (of particular importance for
$\bar{\mbox{H}}$, see Methods), we propose two-photon cooling on $1S \!\!
\rightarrow \!\!3D$ at $205.0 \mbox{ nm}$.  By choosing a comb tooth
spacing of $f_{\mathrm{r}} = 83.5 \mbox{ MHz}$, all six of the allowed
hyperfine and fine structure transitions on $1S \!\!  \rightarrow \!\!
3D$ can be driven simultaneously with a red detuning between
$\gamma/3$ and $\gamma$.  We estimate a resonant excitation rate on
$1S \!\! \rightarrow \!\!  3D$ of $\gamma_{\mathrm{scatt}} \approx
1250 \mbox{ s}^{-1}$ is achievable with demonstrated technology (see
Methods). This excitation rate would produce an acceleration more than
50 times greater than that used in this work to make a MOT of
rubidium.

Atomic oxygen has fine structure in its ${^3}P$ ground state that
spans a range of about $7 \mbox{ THz}$, so the comb's ability to drive
multiple transitions at once is a crucial advantage.  For the
$(2p^4){}^3P \!\!  \rightarrow \!\! (3p){}^3P$ transitions, a
frequency comb centered at $226 \mbox{ nm}$ with a $2 \mbox{ nm}$
bandwidth would be able to drive simultaneous 2-photon transitions for
each fine structure component for a repetition rate near $79.79 \mbox{
  MHz}$. Nitrogen cooling and trapping would proceed on the
$(2p^3){}^4S_{3/2}^{o} \!\!  \rightarrow \!\! (3p){}^4S_{3/2}^{o}$
two-photon transition at $207 \mbox{ nm}$.  Branching to the doublet
manifold limits the total number of quartet excitations per atom to
$\approx 10^3$, sufficient for laser cooling a hot, trapped
sample \cite{HummonPRA08} to the Doppler limit. The hyperfine structure
in the ground state of ${}^{15}N$ is split by $29 \mbox{ MHz}$, so
excitation from a single two-photon tooth may be enough to produce
both cooling and (off-resonant) hyperfine repumping.  Carbon would
likely require multiple combs \cite{Kielpinski:2006ai}, but each would
operate on the same principles we have investigated here.

This demonstration with rubidium confirms the essential aspects of
laser cooling and trapping with frequency combs on 2-photon
transitions.  Future work in extending this technique into the deep UV
should be possible with the addition of frequency conversion stages
for the ML light. In particular, as higher power UV frequency combs
become available \cite{KanaiOE09,ProninNatComm15}, the technology for
laser cooling and trapping will extend the reach of these techniques
to species that cannot currently be produced in ultracold form.

\section{methods}
\subsection{Frequency lock for the optical frequency comb}
To tune the ML laser near the resonance condition for the $5S
\!\!\rightarrow\!\!  5D$ two-photon transition in rubidium
(Fig.~\ref{fig:spectrum}), we sample a fraction of the laser power and
send it to a hot Rb vapor cell in a counter-propagating
geometry \cite{ReinhardtPRA10}. Each excitation to the $5{}^2D_{5/2}$
state produces a spontaneously emitted $420 \mbox{ nm}$ photon as part
of a cascade decay 6.5\% of the time
(Fig.~\ref{fig:spectrum}\textbf{c}), which is collected from the pulse
collision volume and monitored with a photon-counting
detector. Fig.~\ref{fig:spectrum}\textbf{a} shows the resulting
Doppler-free spectrum of the 32 allowed two-photon transitions for
both $^{85}\mathrm{Rb}$ and $^{87}\mathrm{Rb}$ from the $5S$ to $5D$
manifolds. Since the bandwidth of the comb ($< 500
\mbox{ GHz}$) is smaller than the detuning from 1-photon resonance
with the $5P$ states ($\Delta/2 \pi > 1 \mbox{ THz}$), the spectrum
repeats itself with a (two-photon sum frequency) period of
$f_{\mathrm{r}}$. We laser cool and trap $^{85}\mathrm{Rb}$ using the
$5{}^2S_{1/2},F_{\mathrm{g}}=3$ to $5{}^2D_{5/2},F_{\mathrm{e}}=5$
``stretch'' transition.

To maintain sufficient laser stability for Doppler cooling and
trapping, we stabilize the ML laser by locking it to an external
cavity.  The free spectral range of the external cavity is pressure
tuned to be an integer multiple ($q=25$) of the ML laser repetition rate to
guarantee that multiple teeth from across the laser spectrum
contribute to the Pound-Drever-Hall error signal used for the lock.  A
piezo-mounted mirror on the external cavity is then used to stabilize
it to the $5{}^2S_{1/2},F_{\mathrm{g}}=3$ to
$5{}^2D_{5/2},F_{\mathrm{e}}=5$ line using FM spectroscopy of the
vapor cell.  We note that this optical frequency comb is not self
referenced and that we feed back to an unknown combination of
$f_{\mathrm{r}}$ and $f_0$ to maintain the two-photon resonance
condition, which is the only frequency parameter that needs to be
actively stabilized.  The pulse chirp is periodically minimized by
adjusting a Gires-Tournois interferometer in the laser cavity to
maximize the blue light emitted from atoms in the initial CW MOT. The
frequency of the ML laser light used for cooling and trapping is tuned
from the vapor cell lock point using an acousto-optic modulator
downstream.

\subsection{Effective two-photon Rabi frequency}
We model the electric field of a frequency comb propagating along +$z$
in the plane wave approximation as
\begin{eqnarray}
\mathbf{E}(z,t) &=& \frac{\mathcal{E}_{\mathrm{o}}}{2} \mathrm{Env}(t)
\sum_q h(t\!-\!qT_{\mathrm{r}}) \left(
\mbox{\boldmath{$\hat{\epsilon}$}}\,
e^{-i(2 \pi f_{\mathrm{c}}(t-qT_{\mathrm{r}})+ q 2 \pi f_0 T_{\mathrm{r}} - kz)}\right.
\nonumber \\&& \left. \;\;\;\;\;\;\;\;\;\;\; +\,
\mbox{\boldmath{$\hat{\epsilon}$}}^{\ast}
e^{i(2 \pi f_{\mathrm{c}}(t-qT_{\mathrm{r}})+ q 2 \pi f_0 T_{\mathrm{r}} - kz)}
\right) \label{Efield1}
\end{eqnarray}
where $\mathrm{Env}(t)$ is the slowly-varying envelope of the pulse
train (we will take this to be equal to 1),
$\mathcal{E}_{\mathrm{o}}h(0)$ is the peak instantaneous electric
field amplitude, $h(t)$ is the envelope of a single pulse (peaked at
$t=0$), $\mbox{\boldmath{$\hat{\epsilon}$}}$ is the unit vector
describing the laser polarization, $f_{\mathrm{c}}$ is the carrier
frequency of the laser and $f_0$ is the carrier-envelope offset
frequency.  If the pulse envelope $h(t)$ is real and symmetric about
$t=0$ with Fourier transform $\tilde{\mathrm{H}}(\omega)$, we can
rewrite Eq.~(\ref{Efield1}) as
\begin{equation}
\mathbf{E}(z,t) = \sum_p \frac{\mathcal{E}_{p}}{2} \left(
\mbox{\boldmath{$\hat{\epsilon}$}} e^{-i(2 \pi f_p t - kz)} +
\mbox{\boldmath{$\hat{\epsilon}$}}^{\ast} e^{i(2 \pi f_p t - kz)}
\right)
\end{equation}
where $f_p \equiv p f_{\mathrm{r}} + f_0$ is the cyclic frequency of
the $p^{\mathrm{th}}$ optical frequency comb tooth and
\begin{equation}
\mathcal{E}_p = \mathcal{E}_{\mathrm{o}} \frac{\tilde{\mathrm{H}}(2
  \pi (f_p
  - f_{\mathrm{c}}))}{T_{\mathrm{r}}}
\end{equation}
is the time-averaged electric field amplitude of the $p^{\mathrm{th}}$
optical frequency comb tooth.  The single-photon resonant Rabi
frequency for just the $p^{\mathrm{th}}$ tooth to drive the
$\mathrm{g} \! \rightarrow \!\mathrm{i}$ transition is given by
\begin{equation}
g_p^{(\mathrm{gi})} = \frac{e \mathcal{E}_p}{\hbar} \bra{\mathrm{i}}
\mbox{\boldmath{$\hat{\epsilon}$}} \cdot \mathbf{r} \ket{\mathrm{g}}
\end{equation}
(where $\mathbf{r}$ is the position operator for the electron) and
likewise for $\mathrm{i} \! \rightarrow \! \mathrm{e}$, which appear
in Eq.~(\ref{Omegan}) as $g^{(1)}$ and $g^{(2)}$, respectively.

The $5{}^2S_{1/2} \!\! \rightarrow \!\! 5{}^2D_{5/2}$ two-photon
transition near $778 \mbox{ nm}$ in rubidium
primarily gets its strength through single-photon couplings to the
nearby $5{}^2P_{3/2}$ state, which approximation was made implicitly
in Eq.~(\ref{Omegan}).  For the more general case, the $\mathrm{g} \!\!
\rightarrow \!\!\mathrm{e}$ two-photon Rabi frequency includes a sum
over all of the possible intermediate states $\{\ket{\mathrm{i}}\}$.
Matrix elements for calculating AC Stark shifts and photoionization
are similar, but include single photon detunings $\Delta$ for both
emission first and absorption first.  In the case of hydrogen, it is
even important to include continuum states in the sum over
$\mathrm{i}$, which contribute substantially \cite{HaasPRA06}.  The
two-photon resonant Rabi frequency associated with the
$n^{\mathrm{th}}$ tooth of the ``two-photon'' comb can be written
\begin{equation}
\Omega_n =  \sum_{p} \frac{e^2
  \mathcal{E}_{p}\mathcal{E}_{n-p}}{\hbar^2} \bra{\mathrm{e}}
(\mbox{\boldmath{$\hat{\epsilon}$}} \cdot \mathbf{r}) \left(
\sum_{\mathrm{i}}\frac{ \ketbra{\mathrm{i}}{\mathrm{i}}}{2
  \Delta_p^{(\mathrm{i})}} \right) (\mbox{\boldmath{$\hat{\epsilon}$}}
\cdot \mathbf{r}) \ket{\mathrm{g}}.
\end{equation}
For a comb whose spectrum is centered approximately halfway between
$\ket{\mathrm{e}}$ and $\ket{\mathrm{g}}$, the time-averaged laser
intensity $I$ is related to this via $\sum_p \mathcal{E}_p
\mathcal{E}_{n-p} \approx 2 I/ \epsilon_{\mathrm{o}}c$.

In the limit that all of the single-photon detunings
$\Delta_p^{(\mathrm{i})}$ are much larger than the fine and hyperfine
splittings (which is often valid, but is not applicable for the $778
\mbox{ nm}$ line in rubidium), since the term in parentheses includes
a sum over all possible projection quantum numbers, it is rotationally
invariant and the angular momentum prefactors for calculating
$\Omega_n$ arise entirely from the tensor
$(\mbox{\boldmath{$\hat{\epsilon}$}} \cdot
\mathbf{r})(\mbox{\boldmath{$\hat{\epsilon}$}} \cdot \mathbf{r})$.
This limit holds well for hydrogen, and the tensor products can be
used to calculate direct two-photon matrix elements between single
quantum states by using the Wigner-Eckhart theorem and a single
reduced matrix element \cite{HaasPRA06}.  Each irreducible spherical
tensor component contained in $(\mbox{\boldmath{$\hat{\epsilon}$}}
\cdot \mathbf{r})(\mbox{\boldmath{$\hat{\epsilon}$}} \cdot
\mathbf{r})$ can be factored into the product of a
polarization-independent term that contracts $\mathbf{r}$ with itself
and an atom-independent term that depends only on the polarization,
$\mathrm{T}^{(k)}[\mbox{\boldmath{$\hat{\epsilon}$}},
  \mbox{\boldmath{$\hat{\epsilon}$}}]$, which is known as the
polarization tensor \cite{ZareBook}. The rank 0 tensor product
$\mathrm{T}^{(0)}[\mathbf{r},\mathbf{r}]$ is responsible for
${}^2S_{1/2} \!\!  \rightarrow \!\! {}^2S_{1/2}$ transitions, while
the rank 2 tensor $\mathrm{T}^{(2)}[\mathbf{r},\mathbf{r}]$ gives rise
to the $S \!\!  \rightarrow \!\! D$ transition amplitude in analogy to
an electric quadrupole interacting with an electric field gradient.
Reduced matrix elements for two-photon transitions and ionization
rates in hydrogen have been calculated by Haas \textit{et
  al.} \cite{HaasPRA06} for linearly polarized light, and therefore
include the value of the polarization tensor for linear
($\mbox{\boldmath{$\hat{\epsilon}$}}=\hat{\mathbf{z}}$) polarization
$\mathrm{T}^{(k)}[\hat{\mathbf{z}},\hat{\mathbf{z}}]$.

In order to use these reduced matrix elements for the case of
$\sigma^+$ or $\sigma^-$ light, they need to be scaled to reflect the
change of the polarization tensor.  This scaling factor is the term
responsible for the increased strength of the $\sigma^+$ or $\sigma^-$
transitions as compared to $\pi$ polarizations.  For $1S \!\!
\rightarrow \!\! 3D$, there is a convenient comb repetition rate
($f_{\mathrm{r}} = 83.5 \mbox{ MHz}$, see 
Fig.~\ref{fig:HydrogenPredictedSpectrum}) such that all of the
possible fine and hyperfine transitions of $1S \!\!  \rightarrow \!\!
3D$ can be driven simultaneously.  For a single $\sigma^+$ (or
$\sigma^-$) polarized comb on two-photon resonance for $1S \!\!
\rightarrow \!\!  3D$ stretch transitions with unresolved fine and
hyperfine structure, we use the reduced matrix element
$\beta_{ge}^{(2)}$ of Ref.~\cite{HaasPRA06} times the ratio of the
circular to linear rank 2 polarization tensor components
($=\sqrt{6}/2$) to calculate the two-photon resonant Rabi frequency,
\begin{equation}
\Omega/2 \pi = -6.8 \,I  \times 10^{-5} \mbox{ Hz}(\mbox{W/m}^2)^{-1}.
\end{equation}
Likewise, the ionization rate is given by the ionization rate for the
$3D$ state times the fraction of atoms that are in the $3D$ state
($\approx \Omega^2/\gamma^2$):
\begin{equation}
\gamma_{\mathrm{ionization}}/2 \pi = I
\frac{\Omega^2}{\gamma^2}\, 1.9 \times 10^{-6} \mbox{
  Hz}(\mbox{W/m}^2)^{-1}
\end{equation}
where $\gamma/2 \pi = 10.3 \mbox{ MHz}$ is the $3D$ state decay rate
and we have used the reduced matrix elements provided by
Ref.~ \cite{HaasPRA06} rescaled to reflect the value of
$\mathrm{T}^{(2)}_0[\mbox{\boldmath{$\hat{\epsilon}$}},
  \mbox{\boldmath{$\hat{\epsilon}$}}^{\ast}]/\mathrm{T}^{(2)}_0[\hat{\mathbf{z}},
  \hat{\mathbf{z}}]
= -1/2$.

\subsection{Scattering rate from a frequency comb}
To estimate the scattering rate from a frequency comb of coupling
strength $\Omega$ between a ground and excited state (whether it is due to a
single or multi-photon process), we define the resonant saturation
parameter for the $n^{\mathrm{th}}$ comb tooth to be $s_n \equiv
2\Omega_n^2/\gamma^2$, where $\gamma$ is the decay rate of the excited
state, which we will model as decaying only to the ground state.  We
focus on the limit where $s_n \ll 1$ and $\Omega_n T_{\mathrm{r}} \ll
\pi$ due to the low Rabi frequency expected for two-photon
transitions under realistic experimental conditions.  For optical
forces, we are typically most interested in the time-averaged
scattering rate, which permits us to simplify the model by summing up
the scattering rates due to each comb tooth instead of the excitation
amplitudes.  Specifically, the steady-state time-averaged scattering
rate from the $n^{\mathrm{th}}$ comb tooth by a stationary atom will
be given by
\begin{equation}
\gamma_n \approx \gamma \frac{s_n}{2} \frac{1}{1 + (2\delta_n/\gamma)^2}\label{singletooth}
\end{equation}
where $\delta_n \equiv 2 \pi (f_n - f_{\mathrm{ge}})$ is the detuning
of the $n^{\mathrm{th}}$ comb tooth from resonance.  If the center
frequency of the comb of coupling strength is near $f_{\mathrm{ge}}$
and the pulse duration is short compared to the excited state
lifetime, $s_n$ will change very little over the range of $n$ that is
within a few $\gamma$ of resonance and we can approximate $s_n = s_N$
where $N$ is the index of the comb tooth closest to resonance.  In
this case, we can use the identity
\begin{equation}
\sum_{n=-\infty}^{\infty} \frac{1}{1 + a^2(n - b)^2} = \frac{\pi}{a}
\frac{ \mathrm{sinh}\! \left(2 \pi/a  \right)}{ \mathrm{cosh}\!
  \left(2 \pi/a  \right) - \cos(2 \pi b)}
\end{equation}
to write
\begin{equation}
\gamma_{\mathrm{comb}} = \sum_n \gamma_n = \gamma \frac{s_N}{2}
\frac{(\gamma\, T_{\mathrm{r}}/4)\, \mathrm{sinh}\! \left( \gamma\,
  T_{\mathrm{r}}/2 \right)}{ \mathrm{cosh}\! \left( \gamma\,
  T_{\mathrm{r}}/2 \right) - \cos(\delta_NT_{\mathrm{r}})}. \label{sinhcoshMethods}
\end{equation}
In the limit where both $\delta_N/2 \pi$ and $\gamma/2 \pi$ are small
compared to the repetition rate $f_{\mathrm{r}}$, Eq.~(\ref{sinhcoshMethods})
reduces to Eq.~(\ref{singletooth}) with $\gamma_{\mathrm{comb}} \approx
\gamma_N$.  For the laser cooling and trapping we report with
rubidium, the combined effect of all of the off-resonant comb teeth to
the scattering rate when $\delta_N=-\gamma/2$ is approximately
$10^{-4} \gamma_N$, and we can neglect their presence for slow-moving
atoms. For hydrogen laser cooling on $1S \!\!  \rightarrow \!\! 3D$
at $f_{\mathrm{r}}= 83.5 \mbox{ MHz}$, this fraction is less than
0.04, and the single-tooth approximation is likely to be fair.

\subsection{Estimates for application to hydrogen, nitrogen, and oxygen}
\begin{figure}
\begin{center}
\includegraphics[scale=0.6]{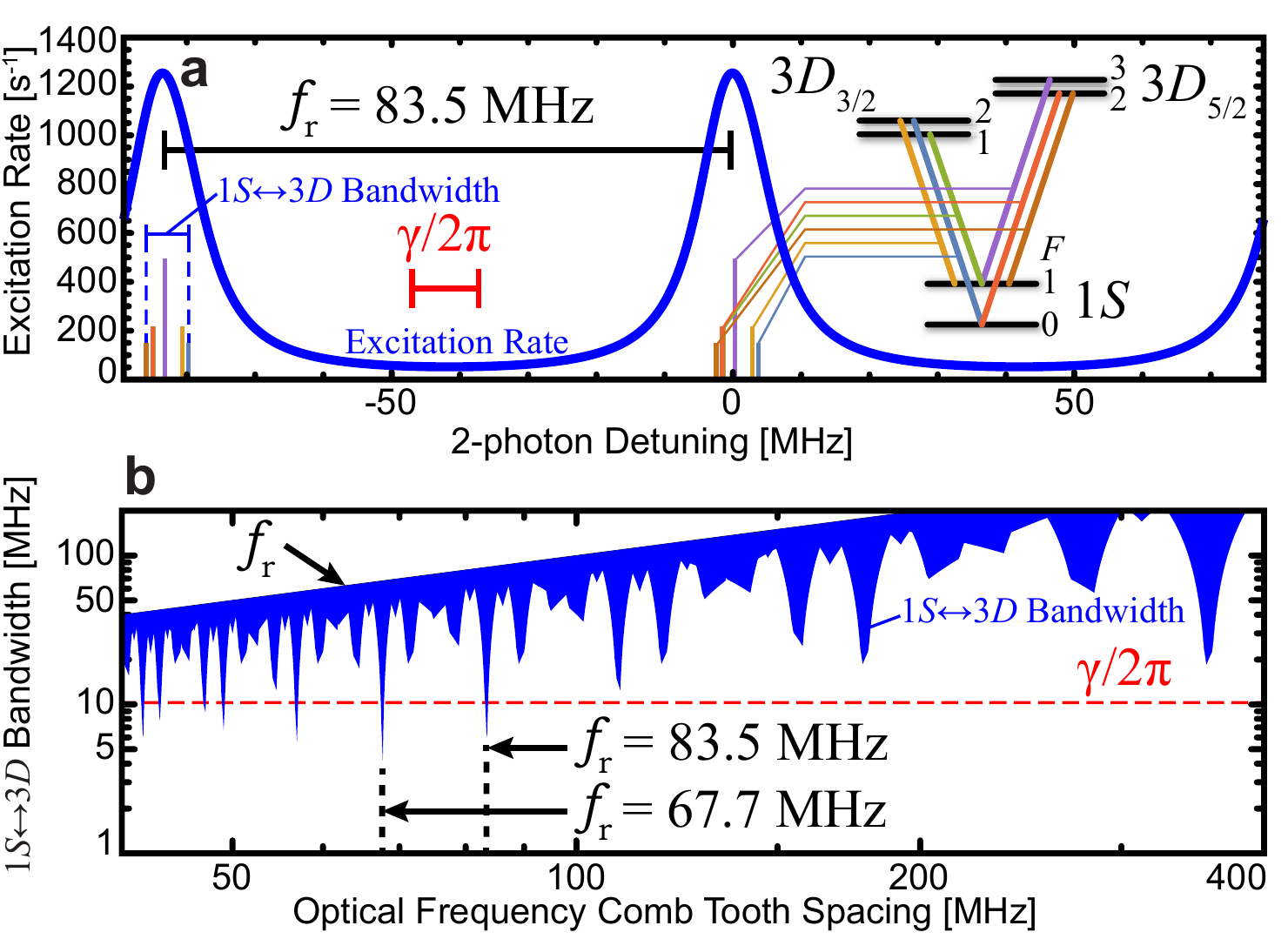}
\end{center}
\caption{Calculated parameters for laser cooling atomic hydrogen on
  $1S \!\! \rightarrow \!\! 3D$.  \textbf{a} Calculated excitation
  rate (blue) as a function of twice the optical frequency (the
  two-photon effective frequency at $102.5 \mbox{ nm}$) for a comb
  with repetition rate $f_{\mathrm{r}} = 83.5 \mbox{ MHz}$.  The
  spectrum shown is the atomic spectrum modulo $f_{\mathrm{r}}$, which
  is how the spectrum will appear when scanning a frequency comb.
  \textbf{b} The bandwidth spanned by the six allowed fine and
  hyperfine transition frequencies $\mathrm{mod} f_{\mathrm{r}}$ is
  less than the natural linewidth of $\gamma/2 \pi = 10.3 \mbox{ MHz}$
  for this repetition rate and a few others.}
\label{fig:HydrogenPredictedSpectrum}
\end{figure}

For H and $\bar{\mbox{H}}$, two-photon Doppler cooling has previously
been proposed on the $1S \!\!  \rightarrow \!\!  2S$ transition
(through forced quenching of the $2S$ state) with a CW
laser \cite{ZehnlePRA01} or optical frequency
comb \cite{Kielpinski:2006ai} centered at $243 \mbox{ nm}$.
Photoionization from the $2S$ state sets a limit on the intensity and
effective (quenched) linewidth for this scheme, which ultimately
limits the scattering rate.  The photoionization cross section of the
$3D$ state is approximately two orders of magnitude smaller than the
$2S$ state for photons at half the state energy \cite{HaasPRA06}.  We
therefore estimate parameters here for two-photon cooling on $1S \!\!
\rightarrow \!\!3D$ at $205.0 \mbox{ nm}$, which is within the phase
matching window for production by frequency doubling in BBO.  We
propose that the added difficulty of producing light at this deeper UV
wavelength is justified by the lower photoionization rate and is
further mitigated by the fact that for this transition, multiple teeth
of the two-photon comb (Eq.~(\ref{Omegan}) and
Fig.~\ref{fig:Schematic}\textbf{c}) can be used simultaneously to
drive different hyperfine and fine-structure transitions in parallel
at no cost in additional laser power
( Fig.~\ref{fig:HydrogenPredictedSpectrum}).  In the limit that both
the average and instantaneous excited state probabilities are small
($\Omega_N \ll \gamma < 2 \pi f_{\mathrm{r}}$) with unequal detunings
from resonance for each transition being driven, coherences between
multiple excited states can be neglected and each line will act
essentially as an independent two-level system.

In Fig.~\ref{fig:HydrogenPredictedSpectrum}, it is shown that that by
choosing a comb tooth spacing of $f_{\mathrm{r}} = 83.5 \mbox{ MHz}$,
all six of the allowed \cite{BoninJOSAB84} hyperfine and fine structure
transitions \cite{KramidaADNDT10} on $1S \!\!  \rightarrow \!\! 3D$ can
be driven simultaneously with a red detuning between $\gamma/3$ and
$\gamma$.  This illustrates the optical frequency comb's ability to
act as its own hyperfine ``repump,'' and allows this scheme to be
applied robustly to magnetically trapped samples, where the presence
of polarization imperfections or off-resonant excitation to undesired
excited states can cause spin flips that must be repumped.  Though not
the focus of this work with two-photon transitions, we have verified
experimentally that we can load and trap rubidium atoms in a
one-photon optical frequency comb MOT that accomplishes its own
hyperfine repumping with another tooth through judicious choice of the
repetition rate.

Optical frequency combs at $205 \mbox{ nm}$ with $\approx 100 \mbox{
  mW}$ of time-averaged power and build-up cavity power enhancement
factors of up to 10 have been
demonstrated \cite{WallIEEEQE03,ZhangOL09,PetersOE09,PetersAnnPhys13}.
Focusing the intra-cavity $1 \mbox{ W}$ beam to a spot size with
diameter $60 \mbox{ }\mu\mbox{m}$ to make a 1D optical molasses would
produce a resonant excitation rate on $1S \!\! \rightarrow \!\!  3D$
that is shown as a function of frequency in 
Fig.~\ref{fig:HydrogenPredictedSpectrum}\textbf{a}. The
photoionization rate at the peak scattering frequency would be
$\gamma_{\mathrm{ionization}} < 0.2 \mbox{ s}^{-1}$ under these
circumstances, so each atom would be able to scatter thousands of
photons before being ionized.

\subsection{Doppler limit for two-photon optical molasses}
To derive the Doppler cooling limit for (approximately equal
frequency) two-photon transitions, we take an algebraic approach to
derive one cooling and two heating mechanisms that will balance one
another in equilibrium \cite{FootAMO}.  We first derive the 1D Doppler
cooling limit for two-photon laser cooling with a CW laser (or a comb
with pulses colliding simultaneously on the atoms), then examine the
situation for an optical frequency comb where there is some finite
delay time that is longer than the pulse duration between forward and
backward propagating pulses.  We will assume that the two-photon
transitions are driven well below saturation (resonant saturation
parameter $s_N \ll 1$) and with a two-photon detuning of $\gamma/2$ to
the red side of resonance. In the case of cooling with an optical
frequency comb, we will assume that the single-tooth approximation
discussed above is valid.

The average cooling force is given by the product of the momentum
transfer per excitation and the excitation rate.  In the limit where
the Doppler shifts are small compared to the excited state linewidth,
the cooling power is given by $\partial E/\partial t
|_{\mathrm{cool}} = -s_N \hbar \omega_{\mathrm{ge}}^2 v^2/c^2$.

This cooling power is balanced by two sources of heating: heating due
to randomly-distributed momentum kicks from absorption events and
heating due to momentum kicks from spontaneous emission \cite{FootAMO}.
For the former, there are only contributions from the single-beam
processes since two-beam absorption does not induce a momentum kick
for counter-propagating beams, and the heating power from absorption
is given by $\partial E/\partial t |_{\mathrm{heat,abs.}} = s_N \gamma \hbar^2
\omega_{\mathrm{ge}}^2/4 m c^2$.

The second heating term is due to spontaneous emission and will depend
upon the details of the decay channels available to the excited state.
If the probability that an excited atom emits a photon with frequency
$\omega_i$ at some point on its way to the ground state is
$\mathcal{P}_i$, the heating from these decays can be modeled with a
probability-weighted sum of the squares of the momentum kicks from
these spontaneously-emitted photons, \textit{viz.}
\begin{equation}
\left. \frac{\partial E}{\partial t} \right|_{\mathrm{heat,spon.}} =
\frac{1}{2m} \gamma_{\mathrm{tot}}\sum_i \mathcal{P}_i \left(\hbar
\frac{\omega_i}{c} \right)^2 \label{SponHeat}
\end{equation} 
where $\gamma_{\mathrm{tot}}$ is the total excitation rate (see
\textit{e.g.} Eq.~(\ref{sinhcoshMethods}) for the case with a single
beam from an optical frequency comb) and we are for the moment
modeling the spontaneous emission as being confined to 1D, which gives
a Doppler limit that agrees with the 3D calculation in the standard
single-photon case.

Eq.~(\ref{SponHeat}) shows the mechanism by which multi-photon cooling
can give rise to a lower Doppler limit than single-photon cooling; by
splitting the decay into smaller, uncorrelated momentum kicks, the
mean square total momentum transfer (and therefore the heating) will
on average be lower than for a single photon decay
channel. Eq.~(\ref{SponHeat}) also shows that there is an additional
heating mechanism for the CW case since $\gamma_{\mathrm{tot}}$ will
in this circumstance include two-beam excitations that are Doppler
free for counter-propagating beams \cite{ZehnlePRA01}.  The excitation
rate from the two-beam terms (which does not contribute to the cooling
in 1D) is 4 times larger than each single-beam term, and the size of
this effect for 1D two-photon laser cooling of atomic hydrogen on a
quenched $1S \!\!  \rightarrow \!\!  2S$ transition, for example,
would lead to a comb-cooled Doppler temperature that is a factor of 2
lower than the predicted CW limit \cite{ZehnlePRA01}.  In order to make
a quantitative estimate of the magnitude of these effects, we model
the decay cascade as proceeding via a single intermediate state
halfway between $\ket{\mathrm{g}}$ and $\ket{\mathrm{e}}$
($\mathcal{P}_i=1$ and $\omega_i = \omega_{\mathrm{ge}}/2$ for
$i=1,2$), which gives us
\begin{equation}
\left. \frac{\partial E}{\partial t} \right|_{\mathrm{heat,spon.}} =
s_N \frac{3 \gamma}{8} \frac{\hbar^2 \omega_{\mathrm{ge}}^2}{m c^2}.\label{CWHeating}
\end{equation}
The equilibrium temperature at which the cooling and heating terms sum
to zero for the CW case gives the Doppler limit for 2-photon 1D
optical molasses with counter-propagating CW laser beams
\begin{equation}
T_{\mathrm{D,CW}} = \frac{5}{4} \frac{\gamma \hbar}{2 k_{\mathrm{B}}}.
\end{equation}
This is 25\% hotter than single-photon cooling on a transition with
the same linewidth, despite the fact that it includes the reduction in
heating from the cascade decay.

For the mode-locked case where pulses from the two directions do not
collide on the atoms at the same time, the cooling power and heating
power from absorption are both the same as the CW case in 1D.
However, the heating power from spontaneous emission is reduced by a
factor of 3 (compare Eq.~(\ref{SponHeatingPower}) and
Eq.~(\ref{CWHeating})) due to the absence of Doppler-free absorption, and
the resulting Doppler cooling limit is given by
Eq.~(\ref{DopplerPrediction}), which is colder than both the CW and the
single-photon cases.

We have extended this model to 3D and performed the detailed decay
channel sum in Eq.~(\ref{SponHeat}) for the rubidium transition used
in this work and find that the calculated Doppler limit of
$T_{\mathrm{D,comb}}=12\mbox{ }\mu\mbox{K}$ agrees well with the
prediction of Eq.~(\ref{DopplerPrediction}).

\subsection{Finite laser tooth linewidth}
By monitoring the $420 \mbox{ nm}$ fluorescence from the pre-cooled
(and then released) rubidium atoms as the ML laser frequency is swept
(shown at the top of Fig.~\ref{fig:molasses}), we obtain a line shape that is
more broad than the natural linewidth of $\gamma/2 \pi = 667 \mbox{
  MHz}$ \cite{ShengPRA08}.  The Doppler broadening expected from motion
would be $630 \mbox{ kHz}$ if taken alone, and the magnetic field is
zeroed to a level where magnetic broadening will not contribute to the
spectral width.  We find that, after taking into account the natural
linewidth and the expected Doppler broadening, we have a residual FWHM
of the two-photon spectrum of around $1.8 \mbox{ MHz}$, which we
attribute to the laser.  It is worth noting that using this width to
infer an optical (that is, single-photon) comb tooth width or vice
versa is highly dependent on the details of the broadening mechanism
(see, \textit{e.g.} \cite{RyanPRA95}), and we therefore rely on the
two-photon spectroscopy exclusively for determining our relevant
effective two-photon spectral linewidth, which is model-independent.
Combining this with the natural linewidth again via convolution gives
us an effective two-photon spectral linewidth with a FWHM of
$\gamma_{\mathrm{eff}}/2 \pi = 2.2 \mbox{ MHz}$.

To account for the effect of finite two-photon spectral linewidth on
scattering rate, we approximate the line shape as Lorentzian to adopt
the model of Haslwanter \textit{et al.} \cite{HaslwanterPRA88}, which
in the low-intensity limit ($s_N \ll 1$) gives the scattering rate
\begin{equation}
\gamma_{N} = \frac{\Omega_N^2}{\gamma}
\frac{\gamma/\gamma_{\mathrm{eff}}}{1 +
  \left(2\delta_N(\mbox{\boldmath$v$})/\gamma_{\mathrm{eff}}
  \right)^2}.\label{SingleToothExcitationRateWithLinewidth}
\end{equation}
We can recognize this as Eq.~(\ref{SingleToothExcitationRate}) with the
replacement
\begin{equation}
\gamma \! \rightarrow \! \gamma_{\mathrm{eff}}, \label{gammaReplacement}
\end{equation}
and conclude that a first approximation of the Doppler temperature
limit can be made in the case of finite spectral linewidth by applying
the replacement Eq.~(\ref{gammaReplacement}) to expressions for the
Doppler temperature (\textit{e.g.} Eq.~(\ref{DopplerPrediction})).
Using this approach for our experimental case where cooling is applied
in 1D but spontaneous emission is approximated as being isotropic in
3D, we predict a Doppler limit of $T_{\mathrm{D,comb}}=31 \mbox{
}\mu\mbox{K}$.

\subsection{Fitting absorption images for temperature}
The spatial width of an atomic cloud following Maxwell-Boltzmann
statistics as a function of time, $t$, is
\begin{equation}
w(t) = \sqrt{w_0^2 + \frac{k_{\mathrm{B}}T}{m} t^2}\label{Widths}
\end{equation}
where $w_0$ is the width at $t=0$ when positions and velocities are
uncorrelated. In Fig.~\ref{fig:molasses} the temperature for data
points labeled as ``Free'' are derived from fitting the free
expansion to Eq.~(\ref{Widths}) where $t=0$ is defined as the end of CW
laser cooling.

For the experiments shown in Fig.~\ref{fig:molasses}, however, we
illuminate the atoms with the ML laser at times $t>0$ which introduces
a damping force.  The simple model of Eq.~(\ref{Widths}) does not
account for the extra dynamics resulting from optical forces.  We
therefore developed a simulation to model an expanding cloud of atoms
(in three dimensions) that is subject to the optical forces of
counterpropogating laser beams in one dimension.  The data points
marked ``Constrained'' in Fig.~\ref{fig:molasses} are derived from
analysis that relies on our simulation.  For each temperature data
point we input experiment parameters (detuning, initial sample
temperature, initial sample width, ML cooling duration, etc.) along
with the experimental measured widths of our atomic cloud during free
expansion.  We run the simulation multiple times as a function of
scattering rate and select the simulation that minimizes $\chi^2$
between the experimentally measured widths and the simulation widths.
From the best simulation we define a temperature using $T =
\frac{m}{k_{\mathrm{B}}}\sigma_v^2$ where $\sigma_v$ is the standard
deviation of the simulation's velocity distribution.  Despite the fact
that the free expansion model does not include effects of the ML
laser, the two methods give almost the same temperatures, which can be
seen by comparing the blue and gray points in Fig.~\ref{fig:molasses}
and the black and red points in the inset of that figure.  There seems
to be a slightly higher inferred temperature when the monte carlo
assisted analysis (``Constrained'') is used in cases where the
acceleration from the ML laser is large.

\begin{acknowledgments}
The authors acknowledge discussions with Chris Monroe and Thomas
Udem, and thank Jun Ye for encouraging them to pursue this work.
The authors thank Anthony Ransford and Anna Wang for technical
assistance, and Andrei Derevianko, Luis Orozco and Trey Porto for
comments on the manuscript. Initial work was supported by the US Air Force Office
of Scientific Research Young Investigator Program under award number
FA9550-13-1-0167, with continuation supported by the NSF CAREER
Program under award number 1455357.  WCC acknowledges support from
the University of California Office of the President's Research
Catalyst Award CA-15-327861.
\end{acknowledgments}

\bibliography{two_photon_cooling}

\end{document}